\documentclass[12pt]{article}
\begin{document}
\newcommand{\beq}{\begin{equation}}
\newcommand{\eeq}{\end{equation}}
\newcommand{\beqa}{\begin{eqnarray}}
\newcommand{\eeqa}{\end{eqnarray}}
\newcommand{\beqar}{\begin{eqnarray*}}
\newcommand{\eeqar}{\end{eqnarray*}}
\newcommand{\al}{\alpha}
\newcommand{\be}{\beta}
\newcommand{\del}{\delta}
\newcommand{\D}{\Delta}
\newcommand{\eps}{\epsilon}
\newcommand{\ga}{\gamma}
\newcommand{\Ga}{\Gamma}
\newcommand{\ka}{\kappa}
\newcommand{\nn}{\nonumber}
\newcommand{\inn}{\!\cdot\!}
\newcommand{\h}{\eta}
\newcommand{\ii}{\iota}
\newcommand{\kk}{\varphi}
\newcommand\F{{}_3F_2}
\newcommand{\la}{\lambda}
\newcommand{\La}{\Lambda}
\newcommand{\na}{\prt}
\newcommand{\Om}{\Omega}
\newcommand{\om}{\omega}
\newcommand{\p}{\phi}
\newcommand{\sig}{\sigma}
\renewcommand{\t}{\theta}
\newcommand{\z}{\zeta}
\newcommand{\ssc}{\scriptscriptstyle}
\newcommand{\eg}{{\it e.g.,}\ }
\newcommand{\ie}{{\it i.e.,}\ }
\newcommand{\labell}[1]{\label{#1}} %{\label{#1}} %
\newcommand{\reef}[1]{(\ref{#1})}
\newcommand\prt{\partial}
\newcommand\veps{\varepsilon}
\newcommand{\pol}{\varepsilon}
\newcommand\vp{\varphi}
\newcommand\ls{\ell_s}
\newcommand\cF{{\cal F}}
\newcommand\cA{{\cal A}}
\newcommand\cS{{\cal S}}
\newcommand\cT{{\cal T}}
\newcommand\cV{{\cal V}}
\newcommand\cL{{\cal L}}
\newcommand\cM{{\cal M}}
\newcommand\cN{{\cal N}}
\newcommand\cG{{\cal G}}
\newcommand\cH{{\cal H}}
\newcommand\cI{{\cal I}}
\newcommand\cJ{{\cal J}}
\newcommand\cl{{\iota}}
\newcommand\cP{{\cal P}}
\newcommand\cQ{{\cal Q}}
\newcommand\cg{{\it g}}
\newcommand\cR{{\cal R}}
\newcommand\CR{{\cal R}}
\newcommand\cB{{\cal B}}
\newcommand\cO{{\cal O}}
\newcommand\tcO{{\tilde {{\cal O}}}}
\newcommand\bg{\bar{g}}
\newcommand\bb{\bar{b}}
\newcommand\bH{\bar{H}}
\newcommand\bX{\bar{X}}
\newcommand\bK{\bar{K}}
\newcommand\bR{\bar{R}}
\newcommand\bZ{\bar{Z}}
\newcommand\bxi{\bar{\xi}}
\newcommand\bphi{\bar{\phi}}
\newcommand\bpsi{\bar{\psi}}
\newcommand\bprt{\bar{\prt}}
\newcommand\bet{\bar{\eta}}
\newcommand\btau{\bar{\tau}}
\newcommand\hF{\hat{F}}
\newcommand\hA{\hat{A}}
\newcommand\hT{\hat{T}}
\newcommand\htau{\hat{\tau}}
\newcommand\hD{\hat{D}}
\newcommand\hf{\hat{f}}
\newcommand\hg{\hat{g}}
\newcommand\hp{\hat{\phi}}
\newcommand\hi{\hat{i}}
\newcommand\ha{\hat{a}}
\newcommand\hb{\hat{b}}
\newcommand\hQ{\hat{Q}}
\newcommand\hP{\hat{\Phi}}
\newcommand\hS{\hat{S}}
\newcommand\hX{\hat{X}}
\newcommand\tL{\tilde{\cal L}}
\newcommand\hL{\hat{\cal L}}
\newcommand\tG{{\widetilde G}}
\newcommand\tg{{\widetilde g}}
\newcommand\tphi{{\widetilde \phi}}
\newcommand\tPhi{{\widetilde \Phi}}
\newcommand\te{{\tilde e}}
\newcommand\tk{{\tilde k}}
\newcommand\tf{{\tilde f}}
\newcommand\ta{{\tilde a}}
\newcommand\tb{{\tilde b}}
\newcommand\tR{{\tilde R}}
\newcommand\teta{{\tilde \eta}}
\newcommand\tF{{\widetilde F}}
\newcommand\tK{{\widetilde K}}
\newcommand\tE{{\widetilde E}}
\newcommand\tpsi{{\tilde \psi}}
\newcommand\tX{{\widetilde X}}
\newcommand\tD{{\widetilde D}}
\newcommand\tO{{\widetilde O}}
\newcommand\tS{{\tilde S}}
\newcommand\tB{{\widetilde B}}
\newcommand\tA{{\widetilde A}}
\newcommand\tT{{\widetilde T}}
\newcommand\tC{{\widetilde C}}
\newcommand\tV{{\widetilde V}}
\newcommand\thF{{\widetilde {\hat {F}}}}
\newcommand\Tr{{\rm Tr}}
\newcommand\tr{{\rm tr}}
\newcommand\STr{{\rm STr}}
\newcommand\hR{\hat{R}}
\newcommand\M[2]{M^{#1}{}_{#2}}

\newcommand\bS{\textbf{ S}}
\newcommand\bI{\textbf{ I}}
\newcommand\bJ{\textbf{ J}}

%\begin{document}
\begin{titlepage}
\begin{center}

\vskip 2 cm
{\LARGE \bf   Higher-derivative couplings and \vskip 0.50 cm
 torsional Riemann curvature
 }\\
\vskip 1.25 cm
   Mohammad R. Garousi\footnote{garousi@um.ac.ir}

\vskip 1 cm
{{\it Department of Physics, Faculty of Science, Ferdowsi University of Mashhad\\}{\it P.O. Box 1436, Mashhad, Iran}\\}
\vskip .1 cm
 \end{center}

\begin{abstract}

Using  the most general higher-derivative field redefinitions for the closed spacetime manifolds, we show that  the tree-level couplings of the metric, $B$-field and dilaton at orders  $\alpha'^2$ and $\alpha'^3$  that have been recently found by the T-duality, can be written in a particular scheme  in terms of the torsional   Riemann curvature $\cR$ and the torsion tensor $H$. The couplings at order $\alpha'^2$ have structures $\cR^3, H^2 \cR^2$, $H^6$, and the couplings at order $\alpha'^3$ have only structures $\cR^4$, $H^2\cR^3$. Replacing $\cR$ with the ordinary Riemann curvature,  the couplings in the structure $H^2\cR^3$ reproduce  the couplings found in the literature by the S-matrix method.
\end{abstract}
\end{titlepage}

\section{Introduction}

String theory is a candidate UV complete theory for quantum gravity. The massless spectrum of the closed string at the critical dimension, $D$, contains graviton,   $B$-field,  and dilaton. The  graviton  corresponds to the diffeomorphism symmetry of the low energy effective action. The diffeomorphism  and the $B$-field gauge transformations have been speculated to combine into   the  generalized diffeomorphism in the double field theory formalism in which the manifest T-duality symmetry $O(D,D)$ is assumed in  the  action  before reduction \cite{Siegel:1993xq,Hull:2009mi,Marques:2015vua}. Even though the generalized geometry appears naturally in the string theory effective action when one reduces the $D$-dimensional theory on tours $T^d$, there are obstacles for the presence of such geometry in string theory before  the Kaluza-Klein (KK) reduction \cite{Hronek:2020xxi}.

The  KK  reduction of the tree-level effective actions of the bosonic and heterotic string theories  on tours $T^d$ have $O(d,d)$ symmetry at all orders of $\alpha'$ \cite{Sen:1991zi,Hohm:2014sxa}. The generalized metric of the internal space appears naturally in the $O(d,d)$ invariant theory. However, in  the KK reduction,  one assumes the reduced fields are independent of the internal tours. Hence, there is no partial derivative  and  no connection made of the generalized metric in the $O(d,d)$ invariant theory, \eg there is no  generalized Riemann curvature  in the internal space. There are, however, partial derivatives and connection  in the base space. Although the metric and $B$-field in the external space can not be combined into a generalized metric, the connection and the torsion in this space may combine to form connection with torsion, \eg there might be  torsional Riemann curvature in the external space.  In fact, there have been  observations from studying the low energy expansion of the  S-matrix elements of  four NS-NS vertex operators at eight-derivative order that  the $D$-dimensional couplings may be in terms of the generalized  Riemann curvature \cite{Gross:1986mw}.

The   gravity couplings at the  eight-derivative order, have been first found   from the sphere-level S-matrix element of four-graviton vertex operators  \cite{ Schwarz:1982jn,Gross:1986iv} as well as from the $\sigma$-model beta function approach \cite{Grisaru:1986vi,Freeman:1986zh}. The result in the Einstein frame for constant dilaton is
\beqa
S&\supset & \frac{\gamma \z(3)}{384\ka^2}\int d^{10}x e^{-3\phi/2} \sqrt{-G}(t_8t_8R^4+\frac{1}{4}\eps_{8}\eps_{8}R^4)\labell{Y0}
\eeqa
where $\gamma=\frac{\alpha'^3}{2^{5}}$ and   $t_8$ is a tensor which is antisymmetric within a pair of indices and is symmetric under exchange of the pair of indices.
The couplings given by $t_8t_8R^4$ have nonzero contribution at four-graviton level, so they were found from the sphere-level S-matrix element of four graviton vertex operators \cite{ Schwarz:1982jn,Gross:1986iv}, whereas the couplings given by $\eps_{8}\eps_{8}R^4$ have nonzero contribution at five-graviton level, \ie $\eps_{8}\eps_{8}R^4$ is total derivative at four-field level. It has been  shown  in \cite{Garousi:2013tca} that $\eps_{8}\eps_{8}R^4$ term is consistent with the sphere-level S-matrix element of five graviton vertex operators. This term contains the Riemann curvature as well as the Ricci and the scalar curvatures.  For the closed spacetime manifolds, one has freedom to use arbitrary higher-derivative field redefinitions \cite{Metsaev:1987zx}. If one uses the field redefinitions that remove all the Ricci and the scalar curvatures, the above action can be written as
\beqa
S&\supset &\frac{\gamma \z(3) }{\ka^2}\int d^{10}x e^{-3\phi/2} \sqrt{-G}\bigg( 2 R_{\alpha}{}^{\epsilon}{}_{\gamma}{}^{\varepsilon} R^{\alpha \beta \gamma \delta} R_{\beta}{}^{\mu}{}_{\epsilon}{}^{\zeta} R_{\delta \zeta \varepsilon \mu} + R_{\alpha \beta}{}^{\epsilon \varepsilon} R^{\alpha \beta \gamma \delta} R_{\gamma}{}^{\mu}{}_{\epsilon}{}^{\zeta} R_{\delta \zeta \varepsilon \mu}\bigg)\labell{RRf}
\eeqa
Using other field redefinitions, the action \reef{Y0}  can be rewritten in various other forms.

 The  B-field and dilaton couplings at four-field level have been added to   \reef{Y0}    by extending the linearized Riemann curvature $\hat{R}$ to the generalized Riemann curvature at the linear order \cite{Gross:1986mw}\footnote{Note that the normalizations of the dilation  and B-field here  are $\sqrt{2}$ and 2 times the normalization of the dilaton and B-field in \cite{Gross:1986mw}, respectively.},
 \beqa
 \bar{R}_{\mu\nu}{}^{\alpha\beta}&= &\hat{R}_{\mu\nu}{}^{\alpha\beta}- \eta_{[\mu}{}^{[\alpha}\phi_{,\nu]}{}^{\beta]}+  e^{-\phi/2}H_{\mu\nu}{}^{[\alpha,\beta]}\labell{trans}
\eeqa
where   the bracket notation is $H_{\mu\nu}{}^{[\alpha,\beta]}=\frac{1}{2}(H_{\mu\nu}{}^{\alpha,\beta}-H_{\mu\nu}{}^{\beta,\alpha})$, and comma denotes the partial derivative.
In the string frame,  the dilaton term  is canceled \cite{Garousi:2012jp}, \ie
\beqa
\bar{R}_{\mu\nu \alpha\beta}&\Longrightarrow&e^{-\phi/2}\bar{\cR}_{\mu\nu \alpha\beta}
\eeqa
where  $\bar{\cR}_{\mu\nu \alpha\beta}$ is the following expression
\beqa
\bar{\cR}_{\mu\nu \alpha\beta}(\Omega)&=&\hat{R}_{\mu\nu \alpha\beta}+H_{\mu\nu [\alpha,\beta]}\labell{RH2}
\eeqa
It is the torsional Riemann curvature  at the linear order, \ie the torsional connection is
\beqa
 {\Omega}^\alpha {}_{\mu\nu}=\Gamma^\alpha{}_{\mu\nu}+\frac{1}{2}H^\alpha{}_{\mu\nu}\labell{linear}
\eeqa
where $\Gamma^\alpha{}_{\mu\nu}$ is the Levi-Civita connection made of the spacetime  metric. The action involving four NS-NS fields  at the sphere level in the string frame is then
\beqa
S\supset \frac{\gamma \z(3)}{384\ka^2} \int d^{10}x e^{-2\phi} \sqrt{-G}(t_8t_8\bar{\cR}^4)\labell{Y3}
\eeqa
While the above action is consistent with the contact terms of the four-point functions, it has been observed in \cite{Garousi:2012yr} that if one replaces \reef{trans} into \reef{RRf}, the resulting four-$H$ couplings are not consistent with the corresponding sphere-level S-matrix element. It means that the  Riemann curvature in the gravity couplings in any arbitrary scheme may not be extended to the torsional Riemann curvature.

If one uses the KK reduction on a circle, then one finds the above couplings are   invariant under the  linearized  T-duality \cite{Garousi:2012jp, Garousi:2013zca}. Requiring the eight-derivative couplings to be invariant under the full T-duality transformations, all NS-NS couplings at order $\alpha'^3$ have been found in \cite{Garousi:2020mqn,Garousi:2020gio} for the closed spacetime manifolds in which one has freedom to use arbitrary field redefinitions. Then it raises  the question that is it possible to write them in terms of the full torsional Riemann curvature?

The natural nonlinear extension of the torsional  Riemann curvature \reef{RH2} is
 \beqa
\cR_{\mu\nu\alpha\beta}(\Omega)&=&R_{\mu\nu\alpha\beta}+H_{\alpha\beta[\mu;\nu]}+\frac{1}{2}H_{\mu[\alpha}{}^ \rho H_{\beta]\nu \rho}\labell{nonlinear2}
%\cR_{ab}&=&R_{ab}+\frac{1}{2}H_{acb}{}_{;c}-\frac{1}{4}H^2_{ab} \,\,\,;\,\,\,\cR=R-\frac{1}{4}H^2\nonumber
\eeqa
where the semicolon denotes the covariant derivative.  It has been observed in \cite{Garousi:2020gio} that there is no scheme in which the NS-NS couplings found in \cite{Garousi:2020gio} can be written in terms of only the nonlinear  generalized Riemann curvature. In fact, the sphere-level couplings of two B-fields and three gravitons at eight-derivative order have been found in \cite{Liu:2019ses} and shown that they can be written in terms of the generalized Riemann curvature and torsion $H$. In this paper, we are going to show that the metric, B-field and dilaton  couplings at  orders $\alpha'^2$, $\alpha'^3$ that have been  found in \cite{Garousi:2019mca,Garousi:2020gio} by T-duality for the closed spacetime manifolds,   can be written in a particular scheme in terms of only generalized Riemann curvature and $H$. The couplings at order $\alpha'$ have been  written in terms of the generalized Riemann curvature and $H$ in \cite{David:2021jqn}.

An outline of the paper is as follows: In section 2 we show that  using the most general  field redefinitions, Bianchi identities and adding total derivative terms, the effective action of the  bosonic string theory at order $\alpha'^2$ can be written in terms of the torsional Riemann curvature and $H$. In section 3, we repeat  the same calculation for the NS-NS couplings of type II superstring theory at order $\alpha'^3$. We show that all couplings can appear in only two structures $\cR^4$ and $H^2\cR^3$.  In subsection 3.1, we show that the coupling in the structure $H^2\cR^3$ can be simplified greatly when the torsional Riemann curvature is replaced by the ordinarily Riemann curvature. We show that the resulting $H^2R^3$ couplings in which the indices of the two $H$ contract with each other at most once, are exactly  the same as the couplings found in \cite{Liu:2019ses} by the S-matrix method.  In section 4, we briefly discuss our results, and write  the couplings at order $\alpha'$ for open spcetime manifolds which has been found in \cite{Garousi:2021yyd} in terms of the  torsional curvatures, $H$ and $\prt\Phi$.

\section{Couplings at order $\alpha'^2$}

In this setion we are going to write the coupligs up to order $\alpha'^2$ in terms of the torsional curvatures. The low energy effective actions of all string theories have the following universal  sector involving the metric, $B$-field and dilaton at the leading order of $\alpha'$ in the string frame:
\beqa
  \bS_0= -\frac{2}{\kappa^2}\int d^{D} x\sqrt{-G} e^{-2\Phi} \left(  R + 4\nabla_{a}\Phi \nabla^{a}\Phi-\frac{1}{12}H^2\right)\labell{S0bf}
\eeqa
where $H$ is field strength of the $B$-field. The KK reduction of this theory on tours $T^d$ has  $O(d,d)$ symmetry \cite{Maharana:1992my}. It can be written in terms of the torsional  scalar curvature as
\beqa
 S_0= -\frac{2}{\kappa^2}\int d^{D} x\sqrt{-G} e^{-2\Phi} \left(  \cR+ 4\nabla_{a}\Phi \nabla^{a}\Phi+\frac{1}{6}H^2\right)\labell{S0bf1}
\eeqa
As can be seen, the  $H$ appears in the connection as torsion and also as a coupling in the  action. This action also includes the first derivative of dilaton. There is no  field redefinition and Bianchi identity at this order, however, one can use integration by parts to rewrite the dilaton coupling in another form as well. So the above  effective action is unique up to a total derivative term.

At the higher orders of $\alpha'$, one should first  use field redefinitions, Bianchi identities and total derivative terms to find the minimum number of independent terms, and then find the coefficients of the independent terms by various techniques in the string theory. It has been shown in \cite{Metsaev:1987zx}, that up to these freedoms,  there are 8 independent basis at order $\alpha'$. The coefficients of the independent terms  have been found in  \cite{Metsaev:1987zx} by the S-matrix method.  The action in a  particular scheme which has no dilaton  is
 \beqa
 \bS_1&=&\frac{-2b_1 }{\kappa^2}\alpha'\int d^{D}x \sqrt{-G}e^{-2\Phi}\Big(   R_{\alpha \beta \gamma \delta} R^{\alpha \beta \gamma \delta} -\frac{1}{2}H_{\alpha}{}^{\delta \epsilon} H^{\alpha \beta \gamma} R_{\beta  \gamma \delta\epsilon}\nn\\
&&\qquad\qquad\qquad\qquad\qquad\quad+\frac{1}{24}H_{\epsilon\delta \zeta}H^{\epsilon}{}_{\alpha}{}^{\beta}H^{\delta}{}_{\beta}{}^{\gamma}H^{\zeta}{}_{\gamma}{}^{\alpha}-\frac{1}{8}H_{\alpha \beta}{}^{\delta} H^{\alpha \beta \gamma} H_{\gamma}{}^{\epsilon \zeta} H_{\delta \epsilon \zeta}\Big)\labell{S1bf}
\eeqa
For the bosonic string theory $b_1=1/4$,   for the heterotic theory  $b_1=1/8$ and for the  superstring theory  $b_1=0$. Note that the Riemann squared term and the first term in the second line are invariant under field redefinitions and total derivative terms, \ie these two terms appear in all other schemes. The other two terms can be written in various other forms in other schemes.  The KK reduction of the above action  on a circle, has  $O(d,d)$ symmetry in a particular scheme in the base space \cite{Kaloper:1997ux,Garousi:2019wgz,Eloy:2020dko}.

Using field redefinitions, Bianchi identities and total derivative terms, one can write the above action in terms of the generalized Riemann curvature and $H$. There are various schemes to write the above action in terms of the generalized Riemann curvature. The couplings in one particular scheme  has been found  in \cite{David:2021jqn}
 \beqa
 S_1=\frac{-2b_1 }{\kappa^2}\alpha'\int d^{D}x \sqrt{-G}e^{-2\Phi}\Big(   \cR_{\alpha \beta \gamma \delta} \cR^{\alpha \beta \gamma \delta} -H_{\alpha}{}^{\delta \epsilon} H^{\alpha \beta \gamma} \cR_{\beta  \gamma \delta\epsilon}-\frac{1}{3}H_{\epsilon\delta \zeta}H^{\epsilon}{}_{\alpha}{}^{\beta}H^{\delta}{}_{\beta}{}^{\gamma}H^{\zeta}{}_{\gamma}{}^{\alpha}\Big)\labell{S1bf1}
\eeqa
Note that when one replaces the generalized Riemann curvature \reef{nonlinear2} into the above equation, one would find no term which has  odd number of $B$-field, and the coefficient of the $H^4$ term becomes the same as the one in \reef{S1bf}.

It has been shown in \cite{Garousi:2019cdn}  that up to field redefinitions, total derivative terms and the Bianchi identities,  there are 60 independent basis at order $\alpha'^2$ in the bosonic string theory. The $O(1,1)$ symmetry  of the circle reduction of these 60 couplings  can fix all parameters up to one overall parameter.  The couplings at order $\alpha'^2$ depends on the couplings at order $\alpha'$ \cite{Bento:1990nv}. The couplings at order $\alpha'^2$ that are correspond to the couplings \reef{S1bf} have been found in \cite{Garousi:2019mca} by the T-duality  in a particular scheme to be
\beqa
\bS_2&=&\frac{-2b_1^2 }{\kappa^2}\alpha'^2\int d^{D}x e^{-2\Phi}\sqrt{-G}\Big[  - \frac{1}{12} H_{\alpha}{}^{\delta \epsilon} H^{\alpha \beta
\gamma} H_{\beta \delta}{}^{\zeta} H_{\gamma}{}^{\iota \kappa}
H_{\epsilon \iota}{}^{\mu} H_{\zeta \kappa \mu}\nn\\&& +
\frac{1}{30} H_{\alpha \beta}{}^{\delta} H^{\alpha \beta
\gamma} H_{\gamma}{}^{\epsilon \zeta} H_{\delta}{}^{\iota
\kappa} H_{\epsilon \zeta}{}^{\mu} H_{\iota \kappa \mu} +
\frac{3}{10} H_{\alpha \beta}{}^{\delta} H^{\alpha \beta
\gamma} H_{\gamma}{}^{\epsilon \zeta} H_{\delta
\epsilon}{}^{\iota} H_{\zeta}{}^{\kappa \mu} H_{\iota \kappa
\mu} \nn\\&&+ \frac{13}{20} H_{\alpha}{}^{\epsilon \zeta}
H_{\beta}{}^{\iota \kappa} H_{\gamma \epsilon \zeta} H_{\delta
\iota \kappa} R^{\alpha \beta \gamma \delta} + \frac{2}{5}
H_{\alpha}{}^{\epsilon \zeta} H_{\beta \epsilon}{}^{\iota}
H_{\gamma \zeta}{}^{\kappa} H_{\delta \iota \kappa} R^{\alpha
\beta \gamma \delta}\nn\\&& + \frac{18}{5} H_{\alpha
\gamma}{}^{\epsilon} H_{\beta}{}^{\zeta \iota} H_{\delta
\zeta}{}^{\kappa} H_{\epsilon \iota \kappa} R^{\alpha \beta
\gamma \delta} -  \frac{43}{5} H_{\alpha \gamma}{}^{\epsilon}
H_{\beta}{}^{\zeta \iota} H_{\delta \epsilon}{}^{\kappa}
H_{\zeta \iota \kappa} R^{\alpha \beta \gamma \delta} \nn\\&&-
\frac{16}{5} H_{\alpha \gamma}{}^{\epsilon} H_{\beta
\delta}{}^{\zeta} H_{\epsilon}{}^{\iota \kappa} H_{\zeta \iota
\kappa} R^{\alpha \beta \gamma \delta} - 2 H_{\beta
\epsilon}{}^{\iota} H_{\delta \zeta \iota}
R_{\alpha}{}^{\epsilon}{}_{\gamma}{}^{\zeta} R^{\alpha \beta
\gamma \delta} - 2 H_{\beta \delta}{}^{\iota} H_{\epsilon
\zeta \iota} R_{\alpha}{}^{\epsilon}{}_{\gamma}{}^{\zeta}
R^{\alpha \beta \gamma \delta}\nn\\&& -  \frac{4}{3}
R_{\alpha}{}^{\epsilon}{}_{\gamma}{}^{\zeta} R^{\alpha \beta
\gamma \delta} R_{\beta \zeta \delta \epsilon} + \frac{4}{3}
R_{\alpha \beta}{}^{\epsilon \zeta} R^{\alpha \beta \gamma
\delta} R_{\gamma \epsilon \delta \zeta} + 3
H_{\beta}{}^{\zeta \iota} H_{\epsilon \zeta \iota} R^{\alpha
\beta \gamma \delta} R_{\gamma}{}^{\epsilon}{}_{\alpha \delta}
\nn\\&&+ 2 H_{\beta \epsilon}{}^{\iota} H_{\delta \zeta \iota}
R^{\alpha \beta \gamma \delta}
R_{\gamma}{}^{\epsilon}{}_{\alpha}{}^{\zeta} + 2 H_{\alpha
\beta \epsilon} H_{\delta \zeta \iota} R^{\alpha \beta \gamma
\delta} R_{\gamma}{}^{\epsilon \zeta \iota} + \frac{13}{10}
H_{\alpha}{}^{\gamma \delta} H_{\beta \gamma}{}^{\epsilon}
H_{\delta}{}^{\zeta \iota} H_{\epsilon \zeta \iota}
\nabla^{\beta}\nabla^{\alpha}\Phi\nn\\&& + \frac{13}{5}
H_{\gamma}{}^{\epsilon \zeta} H_{\delta \epsilon \zeta}
R_{\alpha}{}^{\gamma}{}_{\beta}{}^{\delta}
\nabla^{\beta}\nabla^{\alpha}\Phi -  \frac{52}{5} H_{\beta
\delta}{}^{\zeta} H_{\gamma \epsilon \zeta}
R_{\alpha}{}^{\gamma \delta \epsilon}
\nabla^{\beta}\nabla^{\alpha}\Phi\nn\\&& -  \frac{26}{5} H_{\alpha
\gamma \epsilon} H_{\beta \delta \zeta} R^{\gamma \delta
\epsilon \zeta} \nabla^{\beta}\nabla^{\alpha}\Phi +
\frac{13}{5} \nabla^{\beta}\nabla^{\alpha}\Phi
\nabla_{\epsilon}H_{\beta \gamma \delta}
\nabla^{\epsilon}H_{\alpha}{}^{\gamma \delta} \nn\\&&+ \frac{13}{10}
H_{\beta \gamma}{}^{\epsilon} H^{\beta \gamma \delta}
H_{\delta}{}^{\zeta \iota} \nabla^{\alpha}\Phi
\nabla_{\iota}H_{\alpha \epsilon \zeta}  -  \frac{13}{20}
H_{\alpha}{}^{\beta \gamma} H_{\delta \epsilon}{}^{\iota}
H^{\delta \epsilon \zeta} \nabla^{\alpha}\Phi
\nabla_{\iota}H_{\beta \gamma \zeta} \nn\\&&+ \frac{1}{20}
H_{\alpha}{}^{\delta \epsilon} H^{\alpha \beta \gamma}
\nabla_{\iota}H_{\delta \epsilon \zeta}
\nabla^{\iota}H_{\beta \gamma}{}^{\zeta} + \frac{1}{5}
H_{\alpha}{}^{\delta \epsilon} H^{\alpha \beta \gamma}
\nabla_{\zeta}H_{\gamma \epsilon \iota}
\nabla^{\iota}H_{\beta \delta}{}^{\zeta}\nn\\&& -  \frac{6}{5}
H_{\alpha}{}^{\delta \epsilon} H^{\alpha \beta \gamma}
\nabla_{\iota}H_{\gamma \epsilon \zeta}
\nabla^{\iota}H_{\beta \delta}{}^{\zeta} -  \frac{6}{5}
H_{\alpha \beta}{}^{\delta} H^{\alpha \beta \gamma}
\nabla_{\zeta}H_{\delta \epsilon \iota}
\nabla^{\iota}H_{\gamma}{}^{\epsilon \zeta}\nn\\&&\qquad\qquad\qquad\qquad\qquad\qquad\qquad  + \frac{17}{10}
H_{\alpha \beta}{}^{\delta} H^{\alpha \beta \gamma}
\nabla_{\iota}H_{\delta \epsilon \zeta}
\nabla^{\iota}H_{\gamma}{}^{\epsilon \zeta}\Big]\labell{S2f}
\eeqa
One can add total derivative terms, use field redefinitions and Bianchi identities to write the above couplings in various other schemes. In this section we are going to uses these freedoms to write the couplings in terms of the torsional Riemann curvature and $H$. Since the calculations are lengthy we use package "xAct" \cite{Nutma:2013zea} to perform the calculations in this paper.

In general, consider an action at order $\alpha'^n$,
\beqa
 \bS_n= -\frac{2\alpha'^n}{\kappa^2}\int d^{D} x\sqrt{-G} e^{-2\Phi} \cL_n(G,B,\Phi)\labell{actn}
\eeqa
To add arbitrary  total derivative terms to the above action, we consider the most general total derivative terms at order $\alpha'^n$ in the string frame which have the following structure:
\beqa
-\frac{2\alpha'^n}{\kappa^2}\int d^{D}x \sqrt{-G}e^{-2\Phi} \mathcal{J}_n=-\frac{2\alpha'^n}{\kappa^2}\int d^{D}x\sqrt{-G} \nabla_\alpha (e^{-2\Phi}{\cal I}_n^\alpha) \labell{J3}
\eeqa
where the vector ${\cal I}_n^\alpha$ is   all possible  covariant and gauge invariant  terms at $(2n-1)$-derivative level with even parity.  The coefficient of each term is arbitrary.

The couplings in  $\bS_n$ are also in a  particular  field variables. If one interested in changing only the scheme of the couplings at order $\alpha'^n$, \ie the couplings at orders $\alpha', \cdots, \alpha'^{(n-1)}$ remain fixed,  one can change the field variables in the leading order action \reef{S0bf} as
\begin{eqnarray}
G_{\mu\nu}&\rightarrow &G_{\mu\nu}+\alpha'^n \delta G^{(n)}_{\mu\nu}\nn\\
B_{\mu\nu}&\rightarrow &B_{\mu\nu}+ \alpha'^n\delta B^{(n)}_{\mu\nu}\nn\\
\Phi &\rightarrow &\Phi+ \alpha'^n\delta\Phi^{(n)}\labell{gbp}
\end{eqnarray}
where the tensors $\delta G^{(n)}_{\mu\nu}$, $\delta B^{(n)}_{\mu\nu}$ and $\delta\Phi^{(n)}$ are all possible covariant and gauge invariant terms at $(2n-2)$-derivative level.  $\delta G^{(n)}_{\mu\nu}$, $\delta\Phi^{(n)}$ contain even-parity terms and $\delta B^{(n)}_{\mu\nu}$ contains odd-parity terms.  The above field redefinitions produce the following couplings at order $\alpha'^n$:
\beqa
\delta \!\!\bS_0&\!\!\!\!\!=\!\!\!\!\!\!&\frac{\delta \!\!\bS_0}{\delta G_{\alpha\beta}}\delta G^{(n)}_{\alpha\beta}+\frac{\delta \!\!\bS_0}{\delta B_{\alpha\beta}}\delta B^{(n)}_{\alpha\beta}+\frac{\delta \!\!\bS_0}{\delta \Phi}\delta \Phi^{(n)}
= -\frac{2\alpha'^n}{\kappa^2}\int d^{D} x\sqrt{-G}e^{-2\Phi}\Big[\nn\\&&(\frac{1}{2} \nabla_{\gamma}H^{\alpha \beta \gamma} -  H^{\alpha \beta}{}_{\gamma} \nabla^{\gamma}\Phi)\delta B^{(n)}_{\alpha\beta} -(  R^{\alpha \beta}-\frac{1}{4} H^{\alpha \gamma \delta} H^{\beta}{}_{\gamma \delta}+ 2 \nabla^{\beta}\nabla^{\alpha}\Phi)\delta G^{(n)}_{\alpha\beta}
\nn\\
&&-2( R -\frac{1}{12} H_{\alpha \beta \gamma} H^{\alpha \beta \gamma} + 4 \nabla_{\alpha}\nabla^{\alpha}\Phi -4 \nabla_{\alpha}\Phi \nabla^{\alpha}\Phi)(\delta\Phi^{(n)}-\frac{1}{4}\delta G^{(n)\mu}{}_\mu) \Big]\nn\\
&\equiv &-\frac{2\alpha'^n}{\kappa^2}\int d^{D}x\sqrt{-G}e^{-2\Phi}\mathcal{K}_n
\eeqa
 Adding the total derivative terms and the field redefinition terms  to the action \reef{actn}, one finds new  action $S_n$, \ie
 \beqa
 S_n= -\frac{2\alpha'^n}{\kappa^2}\int d^{D} x\sqrt{-G} e^{-2\Phi} L_n(G,B,\Phi)\labell{S3bf1}
\eeqa
where the  Lagrangian  $L_n(G,B,\Phi)$ is related to the Lagrangian $\cL_n(G,B,\Phi)$ as
\beqa
L_n&=&\cL_n+{\cal J}_n+{\cal K}_n\labell{DLK}
\eeqa
The action $\bS_n$ and $S_n$ are physically equivalent. There is no free parameter in $\cL_n(G,B,\Phi)$. Choosing different values for the arbitrary parameters in ${\cal J}_n$, ${\cal K}_n$, one would find different forms of couplings for the Lagrangian $L_n$.  Alternatively, if one chooses a specific form for the Lagrangian $L_n$ and the above equation has a solution for the arbitrary parameters in ${\cal J}_n$, ${\cal K}_n$, then that Lagrangian would be  physically the same as $\cL_n$. We are looking for the specific  Lagrangian $L_n$ which is in terms of torsional curvature $\cR$,  $H$ and $\prt\Phi$. There are different structures for these couplings. We first consider all possible terms, and then remove some of the structures. If there is a  solution for the above equation, it means the removal of that structures is physically allowed.

To check that the above equation has solution, however,  one should write \reef{DLK}  in terms of independent couplings, \ie    one has to impose the following Bianchi identities:
\beqa
 R_{\alpha[\beta\gamma\delta]}&=&0\nn\\
 \nabla_{[\mu}R_{\alpha\beta]\gamma\delta}&=&0\labell{bian}\\
\nabla_{[\mu}H_{\alpha\beta\gamma]}&=&0\nn\\
{[}\nabla,\nabla{]}O-RO&=&0\nn
\eeqa
 To impose these Bianchi identities in gauge invariant form, one may contract the  left-hand side of each Bianchi identity with the NS-NS field strengths and their derivatives  to produce terms at order $\alpha'^n$. The coefficients of these terms are arbitrary. Adding these terms to the equation \reef{DLK}, then one can check whether or not it has a solution.  Alternatively, to impose the  Bianchi identities in non-gauge invariant form, one may rewrite the terms in \reef{DLK} in  the local frame in which the first derivative of metric is zero, and  rewrite the terms in \reef{DLK} which have derivatives of $H$ in terms of B-field, \ie $H=dB$.  In this way,  the Bianchi identities satisfy automatically \cite{Garousi:2019cdn}.  This latter approach is easier to impose the Bianchi identities by computer. Moreover, in this approach one does not need to introduce another  large number of arbitrary parameters to include the Bianchi identities into the equation \reef{DLK}.

For $n=2$ case, we find that the equation \reef{DLK} has solution if one  removes all couplings involving $\prt\Phi$, and the torsional Ricci and scalar curvatures. The structure $H^4 \cR$ is also allowed to be removed.  We find that not all couplings in the structure $H^6$  can be removed. One couplings in this structure must be in the $L_2$. There are also at least two couplings in the structure $\cR^3$. We then find there are 38 couplings in the structure $H^2\cR^2$.  To write these couplings in terms of independent basis, we first find that there are 25 independent couplings. They are
\beqa
H^2\cO+H^{\alpha\beta\gamma}H^\mu{}_{\beta\gamma}\cO_{\alpha\mu}+H^{\alpha\beta\gamma}H^{\mu\beta}{}_{\gamma}\cO_{\alpha\beta\mu\nu}+H^{\alpha\beta\gamma}H^{\mu\beta\gamma}\cO_{\alpha\beta\gamma\mu\nu\rho}\labell{sss}
\eeqa
where there are 3 basis for $\cO$, 4 basis for $\cO_{\alpha\mu}$, 11 basis for $\cO_{\alpha\beta\mu\nu}$ and 7 basis for $\cO_{\alpha\beta\gamma\mu\nu\rho}$, \ie
\beqa
\cO&=&a_1 \cR_{abcd} \cR^{abcd} + a_2 \cR_{acbd} \cR^{abcd} + a_7 \cR^{abcd} \cR_{cdab}\nn\\
\cO_{\alpha\mu}&=&a_3 \cR_{ab\mu c} \cR_{\alpha }{}^{abc} + a_8 \cR_{bc\mu a} \cR_{\alpha
}{}^{abc} + a_4 \cR_{\alpha }{}^{abc} \cR_{\mu abc} + a_{11} \cR_{\alpha
}{}^{abc} \cR_{\mu bac}\nn\\
\cO_{\alpha\beta\mu\nu}&=& a_{10}
   \cR_{ab\mu  \nu  } \cR_{\alpha  \beta  }{}^{ab} +
   a_{9}
   \cR_{ab\beta  \nu  } \cR_{\alpha  \mu  }{}^{ab} +
   a_{5}
   \cR_{\alpha  }{}^{a}{}_{\mu  }{}^{b} \cR_{\beta  a\nu  b} +
   a_{15}
   \cR_{\alpha  \mu  }{}^{ab} \cR_{\beta  a\nu  b}\nn\\&& +
   a_{13}
   \cR_{\alpha  }{}^{a}{}_{\mu  }{}^{b} \cR_{\beta  b\nu  a} +
   a_{20}
   \cR_{\alpha  \mu  }{}^{ab} \cR_{\beta  \nu  ab} +
   a_{6}
   \cR_{\alpha  }{}^{a}{}_{\beta  }{}^{b} \cR_{\mu  a\nu  b} +
   a_{16}
   \cR_{\alpha  \beta  }{}^{ab} \cR_{\mu  a\nu  b}\nn\\&& +
   a_{14}
   \cR_{\alpha  }{}^{a}{}_{\beta  }{}^{b} \cR_{\mu  b\nu  a} +
   a_{21}
   \cR_{\alpha  \beta  }{}^{ab} \cR_{\mu  \nu  ab} +
   a_{12}
   \cR_{\alpha  }{}^{a}{}_{\mu  }{}^{b} \cR_{\nu  b\beta  a}\nn\\
 \cO_{\alpha\beta\gamma\mu\nu\rho}&=&   a_{19}  \cR_{\alpha  \beta  \mu  }{}^{a} \cR_{\gamma  a\nu
\rho  } +
 a_{24}  \cR_{\alpha  \beta  \mu  }{}^{a} \cR_{\gamma  \nu
\rho  a} +
 a_{22}  \cR_{\alpha  \mu  \beta  }{}^{a} \cR_{\gamma  \nu
\rho  a} +
 a_{25}  \cR_{\alpha  \beta  \gamma  }{}^{a} \cR_{\mu  \nu
\rho  a} \nn\\&&+
 a_{18}  \cR_{\alpha  \beta  \mu  }{}^{a} \cR_{\nu  a\gamma
\rho  } +
 a_{17}  \cR_{\alpha  \mu  \beta  }{}^{a} \cR_{\nu  a\gamma
\rho  } +
 a_{23}  \cR_{\alpha  \beta  \mu  }{}^{a} \cR_{\nu  \rho
\gamma  a}
\eeqa
where $a_1,\cdots, a_{25}$ are some parameters that should be fixed by equating \reef{sss} with the 38 couplings in the structure $H^2\cR^2$.  To write the above basis, we write all contractions of two $H$ and two torsional Riemann curvature $\cR$. To impose the Bianchi identity corresponding to the torsional Riemann curvature, we go to the local frame in which the Levi-Civita connection is zero whereas its derivatives are not zero. Then we find the above independent basis. Note that the above basis have  odd- and even-parity terms. We consider only the even-parity terms of the above basis.

We then equate  the 38 terms to be the same as these basis. To solve the resulting equation, we go to the local frame. We find four parameters $a_{10}$, $a_{11}$, $a_{12}$, $a_{13}$ remain arbitrary. We set them to zero and find 17 non-zero terms. For this particular choices for these parameters, the couplings become
\beqa
S_2&=&\frac{-2b_1^2 }{\kappa^2}\alpha'^2\int d^{D}x e^{-2\Phi}\sqrt{-G}\Big[ -  \frac{4}{3} \cR^{\alpha \beta \gamma
\delta } \cR_{\gamma }{}^{\epsilon }{}_{\alpha }{}^{\varepsilon }
\cR_{\delta \varepsilon \beta \epsilon } + \frac{4}{3} \cR_{\alpha
\gamma }{}^{\epsilon \varepsilon } \cR^{\alpha \beta \gamma
\delta } \cR_{\epsilon \varepsilon \beta \delta }\nn\\&&\quad\qquad\qquad\qquad\qquad\qquad\quad+H^{\alpha\beta\gamma}H^\mu{}_{\beta\gamma}\cO_{\alpha\mu}+H^{\alpha\beta\gamma}H^{\mu\beta}{}_{\gamma}\cO_{\alpha\beta\mu\nu}+
H^{\alpha\beta\gamma}H^{\mu\beta\gamma}\cO_{\alpha\beta\gamma\mu\nu\rho}\nn\\&&
\quad\qquad\qquad\qquad\qquad\qquad\quad- \frac{5}{6} H_{\alpha \beta }{}^{\delta } H^{\alpha \beta
\gamma } H_{\gamma }{}^{\epsilon \varepsilon } H_{\delta
\epsilon }{}^{\zeta } H_{\varepsilon }{}^{\eta \mu } H_{\zeta
\eta \mu } \Big]\labell{aaa}
\eeqa
where the tensors $\cO_{\alpha\mu}, \cO_{\alpha\beta\mu\nu}$ and $\cO_{\alpha\beta\gamma\mu\nu\rho}$ are the following:
\beqa
\cO_{\alpha\mu}&=&-\frac{2}{3} \cR_{ab\mu c} \cR_{\alpha }{}^{abc} + \cR_{bc\mu a} \cR_{\alpha }{}^{abc}-\frac{1}{3} \cR_{\alpha}{}^{abc}\cR_{\mu abc}\nn\\
\cO_{\alpha\beta\mu\nu}&=& \frac{2}{3} \cR_{ab\beta \nu } \cR_{\alpha \mu }{}^{ab} -\frac{8}{3} \cR_{\alpha }{}^{a}{}_{\mu }{}^{b} \cR_{\beta a\nu b} +\frac{8}{3} \cR_{\alpha \mu }{}^{ab} \cR_{\beta a\nu b} -\frac{5}{3} \cR_{\alpha \mu }{}^{ab} \cR_{\beta \nu ab} \nn\\&&-\frac{20}{3} \cR_{\alpha }{}^{a}{}_{\beta }{}^{b} \cR_{\mu a\nu b}+ \frac{20}{3} \cR_{\alpha \beta }{}^{ab} \cR_{\mu a\nu b} -\frac{4}{3} \cR_{\alpha}{}^{a}{}_{\beta}{}^{b}\cR_{\mu b\nu a} - \frac{3}{2} \cR_{\alpha \beta}{}^{ab}\cR_{\mu \nu ab}\nn\\
\cO_{\alpha\beta\gamma\mu\nu\rho}&=&\cR_{\alpha \beta \mu }{}^{a} \cR_{\gamma a\nu \rho } -\frac{22}{3} \cR_{\alpha \beta \mu }{}^{a} \cR_{\gamma \nu \rho a} + \frac{8}{3} \cR_{\alpha \mu \beta }{}^{a} \cR_{\gamma \nu\rho a}\nn\\&& -  \frac{4}{3} \cR_{\alpha \beta \mu }{}^{a} \cR_{\nu a\gamma \rho } -\frac{8}{3} \cR_{\alpha \mu \beta }{}^{a}\cR_{\nu a\gamma \rho }
- 2 \cR_{\alpha \beta \mu }{}^{a}\cR_{\nu\rho\gamma a}
\eeqa
Note that there is no coupling in which all indices of one $H$ contract with all indices of the other $H$, \ie no coupling with structure $H^2\cO$. The couplings \reef{aaa} and the couplings in \reef{S2f} are  the same up to some total derivative terms, field redefinitions and Bianchi identities.

\section{Couplings at order $\alpha'^3$}

It has been shown in \cite{Garousi:2020mqn} that,  up to field redefinitions, total derivative terms and the Bianchi identities,  there are 872 independent basis at order $\alpha'^3$. The $O(1,1)$ symmetry  of the circle reduction of these 872 couplings  fixes all parameters up to one overall parameter. They are \cite{Garousi:2020gio}
\beqa
\bS_3= \frac{\gamma \z(3) }{\ka^2}\int d^{10}x e^{-2\phi} \sqrt{-G}\bigg( 2 R_{\alpha}{}^{\epsilon}{}_{\gamma}{}^{\varepsilon} R^{\alpha \beta \gamma \delta} R_{\beta}{}^{\mu}{}_{\epsilon}{}^{\zeta} R_{\delta \zeta \varepsilon \mu} + R_{\alpha \beta}{}^{\epsilon \varepsilon} R^{\alpha \beta \gamma \delta} R_{\gamma}{}^{\mu}{}_{\epsilon}{}^{\zeta} R_{\delta \zeta \varepsilon \mu}+\cdots\bigg)\labell{RRff}
\eeqa
where dots represent 443  terms that involve $H$ and $\Phi$ (see \cite{Garousi:2020gio} for the explicit form of these couplings). Using field redefinitions, total derivative terms and the Bianchi identities, these terms have been written in \cite{Garousi:2020lof} in terms of 249 couplings that do not involve the dilaton, the Ricci and scalar curvatures. In this section we are going to write the 445 couplings above in terms of the torsional Riemann curvature and $H$.

We first consider all couplings involving $\cR$, $H$, $\prt\Phi$. We do not try to find the independent basis for them at this point. There are 2900 such couplings.  We find  that the equation \reef{DLK} has solution. It means it is possible to write the couplings \reef{RRff} in terms of  $\cR$, $H$, $\prt\Phi$. We then remove the structures that have the dilaton, the torsional Ricci  and the scalar curvatures. The equation \reef{DLK} still has solution. We have also removed the couplings in the structures $H^8$, $H^6\cR$ and $H^4\cR^2$. The equation still has solution. So it means the couplings \reef{RRff} can be written in terms of only $\cR^4$ and $H^2\cR^3$ where $\cR$ is the torsional Riemann curvature. We then choose the couplings in the structure $\cR^4$ to be in the following form:
\beqa
S_3&\supset &\frac{\gamma \z(3) }{\ka^2}\int d^{10}x e^{-2\phi} \sqrt{-G}\Big[a(t_8t_8\cR^4+\frac{1}{4}\eps_{8}\eps_{8}\cR^4)\Big]\labell{Y01}
\eeqa
where in the second term we remove the terms in the expansion which have the torsional Ricci and scalar curvatures. The equation \reef{DLK} still has solution, and fixes the overall factor to be $a=1/384$. There remains 489 couplings in the structure $H^2\cR^3$.  There is no coupling in which all indices of one $H$ contracted with all indices of the other $H$. To write  these 489 couplings  in terms of independent basis, we find the basis of $H^2\cR^3$. There are 254 such basis. Equating  the 489 couplings to be the same as these basis, and going to the local frame to find the parameters of the basis, we find 45 parameters to be arbitrary. We set them to zero, and find  189 non-zero couplings.

The couplings that we have found are the following:
\beqa
S_3&= &\frac{\gamma \z(3) }{\ka^2}\int d^{10}x e^{-2\phi} \sqrt{-G}\Big[\frac{1}{384}(t_8t_8\cR^4+\frac{1}{4}\eps_{8}\eps_{8}\cR^4)+H^{\alpha\beta\gamma}H^{\mu}{}_{\beta\gamma}\cQ_{\alpha\mu}\nn\\
&&\qquad\qquad\qquad\qquad\qquad+H^{\alpha\beta\gamma}H^{\mu\nu}{}_\gamma \cQ_{\alpha\beta\mu\nu}+H^{\alpha\beta\gamma}H^{\mu\nu\rho} \cQ_{\alpha\beta\gamma\mu\nu\rho}\Big]\labell{Y02}
\eeqa
where the tensor $\cQ_{\alpha\mu}$ has the following 17 terms:
\beqa
&&- \frac{5}{6} \cR_{ae}{}^{cd} \cR_{cd\mu b} \cR_{\alpha }{}^{aeb} -
\frac{1}{8} \cR_{cd\mu a} \cR_{eb}{}^{cd} \cR_{\alpha }{}^{aeb} +
\frac{1}{3} \cR_{bdac} \cR_{e}{}^{c}{}_{\mu }{}^{d} \cR_{\alpha
}{}^{aeb} + \frac{1}{2} \cR_{a}{}^{bcd} \cR_{bced} \cR_{\alpha
}{}^{a}{}_{\mu }{}^{e}\nn\\&& -  \frac{1}{4} \cR_{a}{}^{bcd} \cR_{cdeb}
\cR_{\alpha }{}^{a}{}_{\mu }{}^{e} -  \frac{5}{24} \cR_{ecbd}
\cR_{\alpha }{}^{aeb} \cR_{\mu a}{}^{cd} + \frac{31}{48} \cR_{cdeb}
\cR^{cd}{}_{\alpha a} \cR_{\mu }{}^{aeb} + \frac{1}{24} \cR_{cdab}
\cR^{cd}{}_{\alpha e} \cR_{\mu }{}^{aeb} \nn\\&&+ \frac{7}{6} \cR_{cd\alpha
b} \cR_{e}{}^{c}{}_{a}{}^{d} \cR_{\mu }{}^{aeb} -  \frac{4}{3}
\cR_{bcad} \cR_{e}{}^{c}{}_{\alpha }{}^{d} \cR_{\mu }{}^{aeb} +
\frac{1}{4} \cR_{ebcd} \cR_{\alpha }{}^{aeb} \cR_{\mu \
}{}^{c}{}_{a}{}^{d} + \frac{1}{2} \cR_{acbd} \cR_{\alpha }{}^{aeb} \cR_{
\mu }{}^{c}{}_{e}{}^{d}\nn\\&& + \frac{1}{2} \cR_{adbc} \cR_{\alpha
}{}^{aeb} \cR_{\mu }{}^{c}{}_{e}{}^{d} -  \cR_{bdac} \cR_{\alpha
}{}^{aeb} \cR_{\mu }{}^{c}{}_{e}{}^{d} -  \frac{43}{24} \cR_{abcd} \cR_{
\alpha }{}^{aeb} \cR_{\mu e}{}^{cd} + \cR_{acbd} \cR_{\alpha }{}^{aeb}
\cR_{\mu e}{}^{cd} \nn\\&&-  \frac{1}{3} \cR_{bcad} \cR_{\alpha }{}^{aeb}
\cR_{\mu e}{}^{cd}
\eeqa
The tensor $\cQ_{\alpha\beta\mu\nu}$ has the following 81 terms:
\beqa
&& \frac{23}{24} \cR_{bc}{}^{da} \cR_{da\beta \nu } \cR_{\alpha
\mu }{}^{bc}- \frac{7}{24} \cR_{bc}{}^{da} \cR_{da\mu \nu } \cR_{\alpha \beta
}{}^{bc} -  \frac{1}{32} \cR_{bcda} \cR^{bcda} \cR_{\alpha \beta \mu
\nu } -  \frac{1}{16} \cR_{bdca} \cR^{bcda} \cR_{\alpha \beta \mu
\nu } \nn\\&& -  \frac{1}{8} \cR_{bdca} \cR^{bcda} \cR_{\alpha \mu
\beta \nu } -  \frac{1}{16} \cR^{bcda} \cR_{dabc} \cR_{\alpha \mu
\beta \nu } + \frac{7}{3} \cR_{cd\mu a} \cR_{\alpha }{}^{bcd}
\cR_{\beta }{}^{a}{}_{\nu b} - 2 \cR_{ba\mu d} \cR_{\alpha }{}^{bcd}
\cR_{\beta c\nu }{}^{a} \nn\\&&+ 2 \cR_{bd\mu a} \cR_{\alpha }{}^{bcd}
\cR_{\beta c\nu }{}^{a} - 4 \cR_{da\mu b} \cR_{\alpha }{}^{bcd}
\cR_{\beta c\nu }{}^{a} - 3 \cR_{bacd} \cR_{\alpha }{}^{b}{}_{\mu
}{}^{c} \cR_{\beta }{}^{d}{}_{\nu }{}^{a} -  \frac{3}{2} \cR_{bdca}
\cR_{\alpha }{}^{b}{}_{\mu }{}^{c} \cR_{\beta }{}^{d}{}_{\nu }{}^{a}
\nn\\&&+ \frac{7}{6} \cR_{cabd} \cR_{\alpha }{}^{b}{}_{\mu }{}^{c} \cR_{\beta
}{}^{d}{}_{\nu }{}^{a} + \cR_{bcda} \cR_{\alpha }{}^{bcd} \cR_{\beta
\mu \nu }{}^{a} -  \frac{1}{2} \cR_{cdba} \cR_{\alpha }{}^{bcd}
\cR_{\beta \mu \nu }{}^{a} + \cR_{b}{}^{cda} \cR_{da\alpha c} \cR_{\beta
\mu \nu }{}^{b}\nn\\&& + \frac{2}{3} \cR_{ca\mu d} \cR_{\alpha }{}^{bcd}
\cR_{\beta \nu b}{}^{a} -  \frac{5}{6} \cR_{cd\mu a} \cR_{\alpha
}{}^{bcd} \cR_{\beta \nu b}{}^{a} -  \cR_{b}{}^{d}{}_{\alpha }{}^{a}
\cR_{ca\mu d} \cR_{\beta \nu }{}^{bc} -  \frac{19}{6}
\cR_{b}{}^{d}{}_{\alpha }{}^{a} \cR_{cd\mu a} \cR_{\beta \nu }{}^{bc} \nn\\&&-
 \frac{2}{3} \cR_{b}{}^{d}{}_{c}{}^{a} \cR_{da\alpha \mu } \cR_{\beta
\nu }{}^{bc} - 5 \cR_{b}{}^{d}{}_{\alpha }{}^{a} \cR_{da\mu c}
\cR_{\beta \nu }{}^{bc} -  \frac{5}{4} \cR_{da\mu c}
\cR^{da}{}_{\alpha b} \cR_{\beta \nu }{}^{bc} + \frac{25}{24}
\cR_{dabc} \cR^{da}{}_{\alpha \mu } \cR_{\beta \nu }{}^{bc} \nn\\&&+ 2
\cR_{bd\mu a} \cR_{\alpha }{}^{bcd} \cR_{\beta \nu c}{}^{a} +
\frac{4}{3} \cR_{bcda} \cR_{\alpha }{}^{b}{}_{\mu }{}^{c} \cR_{\beta
\nu }{}^{da} -  \frac{16}{3} \cR_{bdca} \cR_{\alpha }{}^{b}{}_{\mu
}{}^{c} \cR_{\beta \nu }{}^{da} + \frac{14}{3} \cR_{cdba} \cR_{\alpha
}{}^{b}{}_{\mu }{}^{c} \cR_{\beta \nu }{}^{da} \nn\\&&-  \frac{13}{8}
\cR_{bcda} \cR_{\alpha \mu }{}^{bc} \cR_{\beta \nu }{}^{da} +
\frac{5}{12} \cR_{bdca} \cR_{\alpha \mu }{}^{bc} \cR_{\beta \nu
}{}^{da} + \cR_{da\alpha b} \cR_{\beta }{}^{bcd} \cR_{\mu
}{}^{a}{}_{\nu c} + \frac{23}{3} \cR_{\alpha }{}^{bcd} \cR_{\beta
c\nu }{}^{a} \cR_{\mu bda}\nn\\&& -  \frac{31}{12} \cR_{\alpha }{}^{bcd}
\cR_{\beta \nu c}{}^{a} \cR_{\mu bda} + \frac{7}{6} \cR_{\alpha
}{}^{bcd} \cR_{\beta \nu b}{}^{a} \cR_{\mu cda} + 3 \cR_{ba\beta d} \cR_{
\alpha }{}^{bcd} \cR_{\mu c\nu }{}^{a} -  \cR_{da\beta b} \cR_{\alpha
}{}^{bcd} \cR_{\mu c\nu }{}^{a} \nn\\&&-  \frac{10}{3} \cR_{\alpha
}{}^{bcd} \cR_{\beta c\nu }{}^{a} \cR_{\mu dba} + \frac{29}{12}
\cR_{\alpha }{}^{bcd} \cR_{\beta \nu c}{}^{a} \cR_{\mu dba} +
\frac{17}{3} \cR_{\alpha }{}^{bcd} \cR_{\beta bc}{}^{a} \cR_{\mu d\nu
a} + 2 \cR_{\alpha }{}^{bcd} \cR_{\beta cb}{}^{a} \cR_{\mu d\nu a} \nn\\&&- 6
\cR_{ca\alpha d} \cR_{\beta \nu }{}^{bc} \cR_{\mu }{}^{d}{}_{b}{}^{a} -
 \frac{10}{3} \cR_{cd\alpha a} \cR_{\beta \nu }{}^{bc} \cR_{\mu
}{}^{d}{}_{b}{}^{a} + \frac{1}{3} \cR_{bacd} \cR_{\alpha
}{}^{b}{}_{\beta }{}^{c} \cR_{\mu }{}^{d}{}_{\nu }{}^{a} +
\frac{3}{2} \cR_{bdca} \cR_{\alpha }{}^{b}{}_{\beta }{}^{c} \cR_{\mu
}{}^{d}{}_{\nu }{}^{a}\nn\\&& -  \frac{1}{2} \cR_{cabd} \cR_{\alpha
}{}^{b}{}_{\beta }{}^{c} \cR_{\mu }{}^{d}{}_{\nu }{}^{a} +
\frac{1}{4} \cR_{\alpha }{}^{bcd} \cR_{\beta cd}{}^{a} \cR_{\mu \nu
ba} + \frac{1}{2} \cR_{ca\beta d} \cR_{\alpha }{}^{bcd} \cR_{\mu \nu
b}{}^{a} -  \frac{11}{12} \cR_{cd\beta a} \cR_{\alpha }{}^{bcd}
\cR_{\mu \nu b}{}^{a} \nn\\&&+ \frac{5}{6} \cR_{b}{}^{d}{}_{\alpha }{}^{a}
\cR_{ca\beta d} \cR_{\mu \nu }{}^{bc} + \frac{19}{12}
\cR_{b}{}^{d}{}_{\alpha }{}^{a} \cR_{cd\beta a} \cR_{\mu \nu }{}^{bc} -
 \frac{1}{12} \cR_{b}{}^{d}{}_{c}{}^{a} \cR_{da\alpha \beta } \cR_{\mu
\nu }{}^{bc} -  \frac{2}{3} \cR_{b}{}^{d}{}_{\alpha }{}^{a}
\cR_{da\beta c} \cR_{\mu \nu }{}^{bc} \nn\\&&+ \frac{1}{4} \cR_{da\beta c}
\cR^{da}{}_{\alpha b} \cR_{\mu \nu }{}^{bc} -  \frac{1}{16} \cR_{dabc}
\cR^{da}{}_{\alpha \beta } \cR_{\mu \nu }{}^{bc} + \frac{2}{3}
\cR_{bd\beta a} \cR_{\alpha }{}^{bcd} \cR_{\mu \nu c}{}^{a} -
\frac{5}{3} \cR_{ba\alpha d} \cR_{\beta }{}^{bcd} \cR_{\mu \nu
c}{}^{a} \nn\\&&-  \frac{3}{2} \cR_{da\alpha b} \cR_{\beta }{}^{bcd} \cR_{\mu
\nu c}{}^{a} -  \frac{43}{24} \cR_{\alpha }{}^{bcd} \cR_{\beta
bc}{}^{a} \cR_{\mu \nu da} + \frac{29}{24} \cR_{\alpha }{}^{bcd} \cR_{
\beta cb}{}^{a} \cR_{\mu \nu da} -  \frac{17}{12} \cR_{bcda}
\cR_{\alpha }{}^{b}{}_{\beta }{}^{c} \cR_{\mu \nu }{}^{da} \nn\\&&-
\frac{1}{6} \cR_{bdca} \cR_{\alpha }{}^{b}{}_{\beta }{}^{c} \cR_{\mu
\nu }{}^{da} + \frac{13}{16} \cR_{bcda} \cR_{\alpha \beta }{}^{bc}
\cR_{\mu \nu }{}^{da} + \frac{1}{4} \cR_{cdba} \cR_{\alpha }{}^{bcd}
\cR_{\mu \nu \beta }{}^{a} -  \frac{1}{2} \cR_{cabd}
\cR^{cd}{}_{\alpha }{}^{a} \cR_{\mu \nu \beta }{}^{b}\nn\\&& -
\frac{1}{2} \cR_{cdba} \cR^{cd}{}_{\alpha }{}^{a} \cR_{\mu \nu \beta
}{}^{b} + \frac{13}{6} \cR_{cd\mu a} \cR_{\alpha }{}^{bcd} \cR_{\nu
}{}^{a}{}_{\beta b} -  \frac{23}{6} \cR_{\alpha }{}^{b}{}_{\mu
}{}^{c} \cR_{\beta b}{}^{da} \cR_{\nu cda} -  \frac{3}{8} \cR_{\alpha
\mu }{}^{bc} \cR_{\beta b}{}^{da} \cR_{\nu cda}\nn\\&& + 5 \cR_{\alpha
}{}^{bcd} \cR_{\beta b\mu }{}^{a} \cR_{\nu cda} -  \frac{3}{2}
\cR_{\alpha }{}^{b}{}_{\beta }{}^{c} \cR_{\mu b}{}^{da} \cR_{\nu cda} +
\frac{9}{16} \cR_{\alpha \beta }{}^{bc} \cR_{\mu b}{}^{da} \cR_{\nu
cda} -  \frac{10}{3} \cR_{\alpha }{}^{bcd} \cR_{\mu b\beta }{}^{a}
\cR_{\nu cda} \nn\\&&+ \frac{13}{3} \cR_{ba\mu d} \cR_{\alpha }{}^{bcd}
\cR_{\nu c\beta }{}^{a} -  \frac{10}{3} \cR_{bd\mu a} \cR_{\alpha
}{}^{bcd} \cR_{\nu c\beta }{}^{a} - 4 \cR_{da\mu b} \cR_{\alpha
}{}^{bcd} \cR_{\nu c\beta }{}^{a} + 7 \cR_{\alpha }{}^{bcd} \cR_{\mu
bc}{}^{a} \cR_{\nu d\beta a} \nn\\&&-  \frac{10}{3} \cR_{\alpha }{}^{bcd}
\cR_{\mu cb}{}^{a} \cR_{\nu d\beta a} + 5 \cR_{ba\alpha d} \cR_{\beta
}{}^{b}{}_{\mu }{}^{c} \cR_{\nu }{}^{d}{}_{c}{}^{a} + \frac{14}{3}
\cR_{bacd} \cR_{\alpha }{}^{b}{}_{\mu }{}^{c} \cR_{\nu }{}^{d}{}_{\beta
}{}^{a} + \frac{1}{6} \cR_{bdca} \cR_{\alpha }{}^{b}{}_{\mu }{}^{c}
\cR_{\nu }{}^{d}{}_{\beta }{}^{a}\nn\\&& -  \frac{7}{2} \cR_{cabd}
\cR_{\alpha }{}^{b}{}_{\mu }{}^{c} \cR_{\nu }{}^{d}{}_{\beta }{}^{a}
\eeqa
And the tensor  $\cQ_{\alpha\beta\gamma\mu\nu\rho}$ has the following 91 terms:
\beqa
&&\frac{1}{48} \cR_{ca\nu \rho } \cR_{\alpha }{}^{bca} \cR_{\beta
b\gamma \mu } + \frac{1}{2} \cR_{ca\gamma \rho } \cR_{\alpha
}{}^{bca} \cR_{\beta b\mu \nu } + \frac{17}{3} \cR_{ca\gamma \rho
} \cR_{\alpha \mu }{}^{bc} \cR_{\beta b\nu }{}^{a} + \frac{23}{24}
\cR_{ca\mu \nu } \cR^{ca}{}_{\alpha b} \cR_{\beta }{}^{b}{}_{\gamma
\rho } \nn\\&&-  \frac{5}{6} \cR_{ba\mu \nu } \cR_{\alpha }{}^{bca}
\cR_{\beta c\gamma \rho } -  \frac{1}{6} \cR_{ca\mu b} \cR_{\alpha
}{}^{bca} \cR_{\beta \gamma \nu \rho } + \frac{25}{6} \cR_{bc\nu
a} \cR_{\alpha }{}^{b}{}_{\mu }{}^{c} \cR_{\beta \gamma \rho }{}^{a}
+ \frac{31}{32} \cR_{ca\mu \nu } \cR_{\alpha }{}^{bca} \cR_{\beta
\gamma \rho b} \nn\\&& -  \frac{13}{12} \cR_{b}{}^{c}{}_{\alpha }{}^{a}
\cR_{ca\mu \nu } \cR_{\beta \gamma \rho }{}^{b} -  \frac{15}{16}
\cR_{ca\mu \nu } \cR^{ca}{}_{\alpha b} \cR_{\beta \gamma \rho }{}^{b}
-  \frac{1}{4} \cR_{ba\mu \nu } \cR_{\alpha }{}^{bca} \cR_{\beta
\gamma \rho c} + \frac{5}{6} \cR_{ca\mu b} \cR_{\alpha }{}^{bca}
\cR_{\beta \nu \gamma \rho }  \nn\\&&+ \frac{5}{3} \cR_{bc\nu a}
\cR_{\alpha }{}^{b}{}_{\mu }{}^{c} \cR_{\beta \rho \gamma }{}^{a} +
\frac{5}{4} \cR_{ca\mu \nu } \cR_{\alpha }{}^{bca} \cR_{\beta \rho
\gamma b} + \cR_{b}{}^{c}{}_{\alpha }{}^{a} \cR_{ca\mu \nu }
\cR_{\beta \rho \gamma }{}^{b} -  \frac{13}{48} \cR_{ca\mu \nu }
\cR^{ca}{}_{\alpha b} \cR_{\beta \rho \gamma }{}^{b} \nn\\&& + \frac{1}{6}
\cR_{ba\mu \nu } \cR_{\alpha }{}^{bca} \cR_{\beta \rho \gamma c} +
\frac{1}{3} \cR_{\alpha \mu }{}^{bc} \cR_{\beta b\nu }{}^{a}
\cR_{\gamma a\rho c} -  \frac{9}{16} \cR_{ca\nu \rho } \cR_{\alpha
\beta }{}^{bc} \cR_{\gamma b\mu }{}^{a} -  \frac{21}{16}
\cR_{b}{}^{a}{}_{\alpha \beta } \cR_{ca\nu \rho } \cR_{\gamma
}{}^{b}{}_{\mu }{}^{c} \nn\\&& -  \frac{25}{24} \cR_{ca\beta \mu }
\cR^{ca}{}_{\alpha b} \cR_{\gamma }{}^{b}{}_{\nu \rho } -
\frac{4}{3} \cR_{ba\beta \mu } \cR_{\alpha }{}^{bca} \cR_{\gamma
c\nu \rho } + \frac{11}{3} \cR_{\alpha }{}^{b}{}_{\mu }{}^{c} \cR_{
\beta b\nu }{}^{a} \cR_{\gamma c\rho a} - 2 \cR_{\alpha }{}^{bca}
\cR_{\beta b\mu c} \cR_{\gamma \nu \rho a} \nn\\&& + \frac{10}{3}
\cR_{\alpha }{}^{bca} \cR_{\beta c\mu b} \cR_{\gamma \nu \rho a} -
\frac{1}{3} \cR_{ba\alpha c} \cR_{\beta }{}^{b}{}_{\mu }{}^{c}
\cR_{\gamma \nu \rho }{}^{a} + \frac{19}{6} \cR_{ca\beta \mu } \cR_{
\alpha }{}^{bca} \cR_{\gamma \nu \rho b} + \cR_{\alpha }{}^{bca} \cR_{
\beta c\mu a} \cR_{\gamma \nu \rho b}  \nn\\&&-  \frac{4}{3}
\cR_{b}{}^{c}{}_{\alpha }{}^{a} \cR_{ca\beta \mu } \cR_{\gamma \nu
\rho }{}^{b} + \frac{7}{4} \cR_{ca\beta \mu } \cR^{ca}{}_{\alpha b}
\cR_{\gamma \nu \rho }{}^{b} -  \frac{10}{3} \cR_{ba\beta \mu }
\cR_{\alpha }{}^{bca} \cR_{\gamma \nu \rho c} + \frac{13}{3}
\cR_{\alpha }{}^{bca} \cR_{\gamma \nu \rho c} \cR_{\mu a\beta b} \nn\\&& +
\frac{5}{3} \cR_{\alpha }{}^{bca} \cR_{\beta \rho \gamma c} \cR_{\mu
a\nu b} + \cR_{\alpha }{}^{bca} \cR_{\beta \nu \gamma \rho }
\cR_{\mu bca} -  \frac{10}{3} \cR_{\alpha }{}^{bca} \cR_{\gamma \nu
\rho c} \cR_{\mu b\beta a} -  \frac{1}{8} \cR_{ca\nu \rho }
\cR_{\alpha }{}^{bca} \cR_{\mu b\beta \gamma } \nn\\&& + \frac{1}{6}
\cR_{ca\gamma \rho } \cR_{\alpha }{}^{bca} \cR_{\mu b\beta \nu } +
\frac{1}{2} \cR_{ca\nu \rho } \cR_{\alpha \beta }{}^{bc} \cR_{\mu
b\gamma }{}^{a} -  \frac{1}{2} \cR_{\alpha }{}^{bca} \cR_{\beta
\gamma \rho c} \cR_{\mu b\nu a} + \frac{7}{3} \cR_{\alpha
}{}^{bca} \cR_{\beta \rho \gamma c} \cR_{\mu b\nu a}  \nn\\&&+ 3 \cR_{\alpha
}{}^{b}{}_{\beta }{}^{c} \cR_{\gamma c\rho }{}^{a} \cR_{\mu b\nu a}
-  \cR_{ca\gamma \rho } \cR_{\alpha \beta }{}^{bc} \cR_{\mu b\nu
}{}^{a} -  \frac{1}{4} \cR_{b}{}^{a}{}_{\alpha \beta } \cR_{ca\nu
\rho } \cR_{\mu }{}^{b}{}_{\gamma }{}^{c} + \frac{1}{2}
\cR_{b}{}^{a}{}_{\alpha \beta } \cR_{ca\gamma \rho } \cR_{\mu
}{}^{b}{}_{\nu }{}^{c} \nn\\&& -  \frac{1}{3} \cR_{\alpha }{}^{bca}
\cR_{\beta \gamma \nu \rho } \cR_{\mu cba} -  \frac{1}{3}
\cR_{\alpha }{}^{bca} \cR_{\beta \nu \gamma \rho } \cR_{\mu cba} + 2
\cR_{\alpha }{}^{bca} \cR_{\gamma \nu \rho b} \cR_{\mu c\beta a} -
\frac{7}{6} \cR_{\alpha }{}^{bca} \cR_{\beta \gamma \rho b} \cR_{\mu
c\nu a} \nn\\&& -  \frac{5}{3} \cR_{\alpha }{}^{bca} \cR_{\beta \rho
\gamma b} \cR_{\mu c\nu a} -  \frac{14}{3} \cR_{ba\alpha c}
\cR_{\gamma \nu \rho }{}^{b} \cR_{\mu }{}^{c}{}_{\beta }{}^{a} +
\frac{4}{3} \cR_{bc\alpha a} \cR_{\gamma \nu \rho }{}^{b} \cR_{\mu
}{}^{c}{}_{\beta }{}^{a} + \frac{19}{6} \cR_{ba\alpha c} \cR_{\beta
\gamma \rho }{}^{b} \cR_{\mu }{}^{c}{}_{\nu }{}^{a}  \nn\\&&-
\frac{5}{6} \cR_{bc\alpha a} \cR_{\beta \gamma \rho }{}^{b} \cR_{\mu
}{}^{c}{}_{\nu }{}^{a} + \cR_{ba\alpha c} \cR_{\beta \rho \gamma
}{}^{b} \cR_{\mu }{}^{c}{}_{\nu }{}^{a} -  \frac{2}{3} \cR_{bc\alpha
a} \cR_{\beta \rho \gamma }{}^{b} \cR_{\mu }{}^{c}{}_{\nu }{}^{a} +
\frac{1}{4} \cR_{\alpha }{}^{bca} \cR_{\beta b\gamma c} \cR_{\mu \nu
\rho a}  \nn\\&&+ \frac{1}{4} \cR_{\alpha }{}^{bca} \cR_{\beta c\gamma b}
\cR_{\mu \nu \rho a} + \frac{3}{16} \cR_{ca\beta \gamma }
\cR_{\alpha }{}^{bca} \cR_{\mu \nu \rho b} -  \frac{1}{4}
\cR_{\alpha }{}^{bca} \cR_{\beta c\gamma a} \cR_{\mu \nu \rho b} -
\frac{1}{8} \cR_{b}{}^{c}{}_{\alpha }{}^{a} \cR_{ca\beta \gamma }
\cR_{\mu \nu \rho }{}^{b}  \nn\\&&+ \frac{1}{16} \cR_{ca\beta \gamma }
\cR^{ca}{}_{\alpha b} \cR_{\mu \nu \rho }{}^{b} + \frac{1}{8} \cR_{ba
\beta \gamma } \cR_{\alpha }{}^{bca} \cR_{\mu \nu \rho c} -
\frac{19}{24} \cR_{ba\alpha \beta } \cR_{\gamma }{}^{b}{}_{\mu
}{}^{c} \cR_{\nu }{}^{a}{}_{\rho c} -  \frac{1}{2} \cR_{ba\alpha
\beta } \cR_{\mu }{}^{b}{}_{\gamma }{}^{c} \cR_{\nu }{}^{a}{}_{\rho
c}  \nn\\&&-  \frac{11}{3} \cR_{\alpha }{}^{b}{}_{\mu }{}^{c} \cR_{\gamma
c\rho }{}^{a} \cR_{\nu b\beta a} -  \frac{1}{4} \cR_{ca\beta \mu }
\cR^{ca}{}_{\alpha b} \cR_{\nu }{}^{b}{}_{\gamma \rho } -
\frac{1}{3} \cR_{ba\beta \mu } \cR_{\alpha }{}^{bca} \cR_{\nu
c\gamma \rho } - 2 \cR_{\alpha \mu }{}^{bc} \cR_{\beta b\gamma
}{}^{a} \cR_{\nu c\rho a}  \nn\\&&-  \frac{1}{24} \cR_{\alpha \beta
}{}^{bc} \cR_{\gamma b\mu }{}^{a} \cR_{\nu c\rho a} + \frac{1}{3}
\cR_{ba\alpha \beta } \cR_{\gamma }{}^{b}{}_{\mu }{}^{c} \cR_{\nu
\rho c}{}^{a} -  \frac{7}{3} \cR_{\alpha }{}^{bca} \cR_{\beta b\mu
c} \cR_{\nu \rho \gamma a} + \frac{9}{2} \cR_{\alpha }{}^{bca}
\cR_{\beta c\mu b} \cR_{\nu \rho \gamma a} \nn\\&& -  \frac{7}{6}
\cR_{\alpha }{}^{bca} \cR_{\mu b\beta c} \cR_{\nu \rho \gamma a} -
\frac{1}{6} \cR_{\alpha }{}^{bca} \cR_{\mu c\beta b} \cR_{\nu \rho
\gamma a} -  \frac{7}{3} \cR_{bc\beta a} \cR_{\alpha }{}^{b}{}_{\mu
}{}^{c} \cR_{\nu \rho \gamma }{}^{a} + \frac{10}{3} \cR_{ba\alpha
c} \cR_{\beta }{}^{b}{}_{\mu }{}^{c} \cR_{\nu \rho \gamma }{}^{a} \nn\\&& -
\frac{29}{6} \cR_{ca\alpha b} \cR_{\beta }{}^{b}{}_{\mu }{}^{c}
\cR_{\nu \rho \gamma }{}^{a} -  \frac{10}{3} \cR_{ba\alpha c}
\cR_{\mu }{}^{b}{}_{\beta }{}^{c} \cR_{\nu \rho \gamma }{}^{a} +
\frac{5}{3} \cR_{ca\alpha b} \cR_{\mu }{}^{b}{}_{\beta }{}^{c}
\cR_{\nu \rho \gamma }{}^{a} -  \frac{17}{6} \cR_{ca\beta \mu }
\cR_{\alpha }{}^{bca} \cR_{\nu \rho \gamma b}  \nn\\&&-  \frac{5}{6}
\cR_{\alpha }{}^{bca} \cR_{\beta c\mu a} \cR_{\nu \rho \gamma b} -
\frac{13}{6} \cR_{\alpha }{}^{bca} \cR_{\mu c\beta a} \cR_{\nu \rho
\gamma b} -  \frac{1}{6} \cR_{b}{}^{c}{}_{\alpha }{}^{a}
\cR_{ca\beta \mu } \cR_{\nu \rho \gamma }{}^{b} -  \frac{7}{24}
\cR_{ca\beta \mu } \cR^{ca}{}_{\alpha b} \cR_{\nu \rho \gamma }{}^{b}
 \nn\\&&-  \frac{1}{6} \cR_{ba\beta \mu } \cR_{\alpha }{}^{bca} \cR_{\nu
\rho \gamma c} -  \frac{5}{3} \cR_{ba\alpha \mu } \cR_{\beta
}{}^{b}{}_{\nu }{}^{c} \cR_{\rho }{}^{a}{}_{\gamma c} +
\frac{161}{96} \cR_{ca\mu \nu } \cR^{ca}{}_{\alpha b} \cR_{\rho
}{}^{b}{}_{\beta \gamma } -  \frac{7}{3} \cR_{ba\mu \nu }
\cR_{\alpha }{}^{bca} \cR_{\rho c\beta \gamma }  \nn\\&&+ \frac{4}{3}
\cR_{\alpha }{}^{b}{}_{\mu }{}^{c} \cR_{\beta b\nu }{}^{a} \cR_{\rho c
\gamma a} -  \frac{5}{3} \cR_{\alpha }{}^{bca} \cR_{\mu b\beta \nu
} \cR_{\rho c\gamma a} -  \cR_{\alpha }{}^{b}{}_{\beta }{}^{c}
\cR_{\mu b\nu }{}^{a} \cR_{\rho c\gamma a}
\eeqa
  Here also the tensors  $\cQ_{\alpha\mu}$,  $\cQ_{\alpha\beta\mu\nu}$,  and $\cQ_{\alpha\beta\gamma\mu\nu\rho}$ have even and odd parities. We consider only their even-parity parts. The action \reef{Y02} contains only the torsional Riemann curvature and the torsion tensor $H$. It is the same as \reef{RRff} up to field redefinitions, total derivative terms and the Bianchi identities.

We have tried to write the above couplings in terms of the following couplings:
\beqa
t_8t_8H^2\cR^3&=&t_8^{\mu_1\cdots \mu_8}t_8^{\nu_1\cdots \nu_8}H_{\mu_1\mu_2\lambda}H_{\nu_1\nu_2}{}^\lambda\cR_{\mu_3\mu_4\nu_3\nu_4}\cR_{\mu_5\mu_6\nu_5\nu_6}\cR_{\mu_7\mu_8\nu_7\nu_8}\nn\\
\epsilon_9\epsilon_9H^2\cR^3&=&-\epsilon_9^{\mu_1\cdots \mu_8\alpha\lambda}\epsilon_9^{\nu_1\cdots \nu_8\beta}{}_{\lambda}H_{\mu_1\mu_2\beta}H_{\nu_1\nu_2\alpha}\cR_{\mu_3\mu_4\nu_3\nu_4}\cR_{\mu_5\mu_6\nu_5\nu_6}\cR_{\mu_7\mu_8\nu_7\nu_8}\labell{e9e9}
\eeqa
where we must remove the torsional Ricci and scalar curvatures  in the expansion of the second line above because there is no such tensors in \reef{Y02}. One can write the couplings \reef{Y02} in terms of above couplings and  some extra terms as in  $\cQ_{\alpha\mu}$,  $\cQ_{\alpha\beta\mu\nu}$,  and $\cQ_{\alpha\beta\gamma\mu\nu\rho}$. However, we have found that they just change the coefficients of some of the terms in $\cQ_{\alpha\mu}$,  $\cQ_{\alpha\beta\mu\nu}$,  and $\cQ_{\alpha\beta\gamma\mu\nu\rho}$. One also need to include some other couplings which have the structure $H^2\cQ$. Since that form of the couplings is not illuminating, we did not write  \reef{Y02} in that form. The  couplings \reef{e9e9}  have been introduced in \cite{Liu:2013dna,Grimm:2017okk} to write the couplings of two $B$-field and three gravitons found by the S-matrix elements, in terms of these tensors. It has been  observed in \cite{Liu:2019ses} that the tree-level couplings can not be written in terms of only these tensors. In the next subsection, we consider the two $B$-field and three graviton couplings in the action \reef{Y02}.

\subsection{$H^2R^3$ couplings}

If one is  interested only in the gravity part of the torsional Riemann curvature, the results simplify greatly. In fact there are only 24 basis for $H^2R^4$. The are 2 basis in $Q$, 3 basis in $Q_{\alpha\mu}$, 11 basis in  $Q_{\alpha\beta\mu\nu}$ and 8 basis in  $Q_{\alpha\beta\gamma\mu\nu\rho}$. One can write the couplings in terms of these 24 independent basis. However, to compare the result with the couplings found in \cite{Liu:2019ses}, we write the couplings in terms of $t_8t_8H^2R^3$, $\epsilon_9\epsilon_9H^2R^3$ and the basis $Q$, $Q_{\alpha\mu}$,  $Q_{\alpha\beta\mu\nu}$ and $Q_{\alpha\beta\gamma\mu\nu\rho}$.

We write the action \reef{Y02} in the following form:
\beqa
S_3&\supset &\frac{\gamma \z(3) }{\ka^2}\int d^{10}x e^{-2\phi} \sqrt{-G}\Big[\frac{1}{384}(t_8t_8\cR^4+\frac{1}{4}\eps_{8}\eps_{8}\cR^4)+a t_8t_8H^2R^3+b \epsilon_9\epsilon_9H^2R^3\labell{Y02G}\\
&&\qquad\qquad\qquad\qquad+H^2 Q+H^{\alpha\beta\gamma}H^{\mu}{}_{\beta\gamma}Q_{\alpha\mu}+H^{\alpha\beta\gamma}H^{\mu\nu}{}_\gamma Q_{\alpha\beta\mu\nu}+H^{\alpha\beta\gamma}H^{\mu\nu\rho} Q_{\alpha\beta\gamma\mu\nu\rho}\Big]\nn
\eeqa
where $a,b$ are two parameters. Note that the curvatures in $t_8t_8\cR^4$ and $\eps_{8}\eps_{8}\cR^4$  is the torsional Riemann curvature, whereas in all other terms the curvature is the standard Riemann curvature. Note that in the expansion of  $\eps_{8}\eps_{8}$ and $\eps_{9}\eps_{9}$ we must remove the Ricci and scalar curvatures  because there is no such tems in \reef{Y02}. Equating the above couplings with the two $H^2R^3$ terms in \reef{Y02}, and going to the local frame to impose the Bianchi identity,  one finds
\beqa
a\,=\,-\frac{1}{192}&;& b\,=\,- \frac{1}{2304}
\eeqa
The tensors $Q_{\alpha\mu}$,  $Q_{\alpha\beta\mu\nu}$ are zero, and
\beqa
Q&=&- \frac{1}{18} R_{a}{}^{e}{}_{c}{}^{f} R^{abcd} R_{bfde} +
\frac{1}{72} R_{ab}{}^{ef} R^{abcd} R_{cdef}\nn\\
Q_{\alpha\beta\gamma\mu\nu\rho}&=& \frac{1}{2} R_{\alpha \beta }{}^{bc} R_{\mu abc} R_{\nu
\rho \gamma }{}^{a} +  \frac{1}{4} R_{\gamma }{}^{abc} R_{\mu
\nu \alpha \beta } R_{\rho abc} - \frac{1}{2} R_{\alpha \beta
b}{}^{c} R_{\mu \nu a}{}^{b} R_{\rho c\gamma }{}^{a} +
\frac{2}{3} R_{\mu a\alpha }{}^{b} R_{\nu b\beta }{}^{c}
R_{\rho c\gamma }{}^{a} \nn\\&&+2 R_{\mu \alpha a}{}^{b} R_{\nu \beta
b}{}^{c} R_{\rho c\gamma }{}^{a} + 2 R_{\mu abc} R_{\nu \alpha
}{}^{bc} R_{\rho \beta \gamma }{}^{a} - 4 R_{\mu abc} R_{\nu
}{}^{a}{}_{\alpha }{}^{c} R_{\rho \beta \gamma }{}^{b}
\eeqa
Apart from the couplings $H^2Q$, the above results are exactly the couplings that have been found in \cite{Liu:2019ses} by the tree-level five-point functions including the correct normalization of the couplings in the structure $H^{\alpha\beta\gamma}H^{\mu\nu\rho} Q_{\alpha\beta\gamma\mu\nu\rho}$ that have been clarified in \cite{Liu:2022bfg}. The presence of these terms have been also confirmed by T-duality in \cite{Wulff:2021fhr,David:2021jqn}. Note that the five-point S-matrix calculations can not fix the Ricci and scalar curvatures, hence, one should also remove these terms in the expansion of $\epsilon_9\epsilon_9H^2R^3$ term in \cite{Liu:2019ses}.

In comparing the results in \cite{Liu:2019ses} with the couplings in \reef{Y02G} we did not consider the couplings in the expansion of $\eps_{8}\eps_{8}\cR^4$ which
involve torsional Ricci and scalar curvatures. While there are no such terms in \reef{Y02G}, the couplings in \cite{Liu:2019ses} include such terms. If one replaces the expression \reef{nonlinear2} into those terms, one would find the couplings in \reef{Y02G} as well as the following  couplings:
\beqa
H^2Q&=& H^2\Big(-\frac{1}{12} R_{a}{}^{e}{}_{c}{}^{f} R^{abcd} R_{bfde} +
\frac{1}{48} R_{ab}{}^{ef} R^{abcd} R_{cdef}\Big)\\
H^{\alpha\beta\gamma}H^{\mu}{}_{\beta\gamma}Q_{\alpha\mu}&=&H^{\alpha\beta\gamma}H^{\mu}{}_{\beta\gamma}\Big(
R_{bfde} R_{\alpha }{}^{e}{}_{c}{}^{f} R_{\mu }{}^{bcd}+
\frac{1}{2} R_{b}{}^{def} R_{cdef} R_{\mu }{}^{b}{}_{\alpha
}{}^{c}+\frac{1}{4} R_{cdef} R_{\alpha b}{}^{ef} R_{\mu }{}^{bcd}\Big)\nn
\eeqa
Hence, the above couplings are the difference between the couplings found in \cite{Liu:2019ses} and the couplings \reef{Y02G} that are produced by the T-duality. On the other hand, using equations of motion
\beqa
&&0=R +4\nabla_\mu\nabla^\mu \phi -4\nabla_\mu\phi\nabla^\mu \phi-\frac{1}{12}H^{\mu\nu\rho}H_{\mu\nu\rho}\nn\\
&&0=R_{\mu\nu} +2\nabla_\mu\nabla_\nu \phi-\frac{1}{4} H_{\mu}^{\rho\sigma}H_{\nu\rho\sigma}\labell{eom}
\eeqa
and using the fact that the contact terms of the S-matrix elements are zero for the Ricci and scalar curvatures,  one can write $H^2$ and $H^{\alpha\beta\gamma}H^{\mu}{}_{\beta\gamma}$ in terms of dilaton. Hence, up to equations of motion, the difference  between the couplings found in \cite{Liu:2019ses} and the couplings \reef{Y02G} are some dilaton couplings. In \cite{Liu:2019ses}, the dilaton is assumed to be constant.

For non-constant  dilaton, we expect the S-matrix calculation should reproduce the couplings in \reef{Y02G}. In particular, the non-constant dilaton should produce the $H^2Q$ terms in \reef{Y02G}. These terms
 might be resulted from  residual contact terms in comparing the massless poles of the field theory including the dilaton pole,  and the massless poles of the sphere-level S-matrix element of two $B$-field and three graviton vertex operators that did not considered in \cite{Liu:2019ses}.  Even though there are no such residual contact terms for constant dilaton, there might be such contact terms for non-constant dilaton. %two $B$-field and three graviton amplitude. %  %Since these terms are resulting from the torsional Ricci and scalar curvatures, the S-matrix calculations could not fix them.

\section{Discussion}

In this paper we have shown that  the effective action of the bosonic string theory at order $\alpha'^2$, and the NS-NS couplings of the type II superstring theory at order $\alpha'^3$ that have been found in \cite{Garousi:2019mca,Garousi:2020gio}  by the T-duality, can be written in terms of the torsional Riemann curvature and the torsion tensor $H$, \ie equations \reef{aaa} and \reef{Y02}. In this study we have used the most general field redefinitions, Bianchi identities and integration by parts. The arbitrary field redefinitions  are allowed only for the spacetime manifolds which have  no boundary. Hence, the couplings \reef{S0bf1}, \reef{aaa} and \reef{Y02} are not background independent. They are valid only for the background in which the spacetime is a closed manifold. To find background independent couplings, one must find the couplings for the background in which the spacetime is an open manifold.  We expect the background independent effective actions to be in terms of the torsional Riemann, Ricci and scalar curvatures, as well as the torsion $H$ and $\prt\Phi$.

In the presence of boundary,  one is not allowed to use the arbitrary field redefinitions because the field redefinitions must respect the information on the boundary \cite{Garousi:2021cfc}. In this case,  one may still use some restricted field redefinitions \cite{Garousi:2021yyd}. The effective action for the open spacetime manifold, can not be found by using the restricted field redefinitions on the couplings \reef{S0bf1}, \reef{aaa} and \reef{Y02} because these couplings  are found for the closed spacetime manifolds. Note that  the reverse is true, \ie if one somehow finds the effective action for open spacetime manifolds, it would be also valid for the closed spacetime manifold as well. One may also use the most general field redefinitions which are allowed for the closed spacetime manifold,  to simplify the action for the closed spacetime manifolds, \eg the actions  \reef{S0bf1}, \reef{aaa} and \reef{Y02}.

To find the effective action for open spacetime manifolds,  one may use the restricted field redefinitions to find all independent couplings involving the metric, $B$-field and the  dilaton. Then one should reduce them on a circle and impose the $O(1,1)$ symmetry to fix  the parameters of the independent couplings. In this way, however, the $O(1,1)$ symmetry can not fix all independent couplings. One needs also to use the cosmological reduction and impose the $O(d,d)$ symmetry for fixing all parameters \cite{Garousi:2021yyd}.  The result for the couplings at order $\alpha'$ has been found in \cite{Garousi:2021yyd} to be
\beqa
\bS_1
&=&-\frac{2b_1}{\kappa^2} \alpha'\int d^{26}x \sqrt{-G} e^{-2\Phi}\Bigg[R^2_{\rm GB}+\frac{1}{24} H_{\alpha }{}^{\delta \epsilon } H^{\alpha \beta
\gamma } H_{\beta \delta }{}^{\varepsilon } H_{\gamma \epsilon
\varepsilon }-\frac{1}{8}  H_{\alpha \beta }{}^{\delta }
H^{\alpha \beta \gamma } H_{\gamma }{}^{\epsilon \varepsilon }
H_{\delta \epsilon \varepsilon }\nn\\&&\qquad\qquad +
R^{\alpha \beta }H_{\alpha }{}^{\gamma \delta } H_{\beta \gamma \delta }  -\frac{1}{12} R H_{\alpha
\beta \gamma } H^{\alpha \beta \gamma }  -\frac{1}{2} H_{\alpha }{}^{\delta \epsilon } H^{
\alpha \beta \gamma } R_{\beta \gamma \delta \epsilon
}\nn\\&&\qquad\qquad +4R \nabla_{\alpha }\Phi
\nabla^{\alpha }\Phi -16
 R^{\alpha \beta }\nabla_{\alpha }\Phi \nabla_{\beta
}\Phi \Bigg]\labell{ffinal}
\eeqa
where $R^2_{\rm GB}$ is the Gauss-Bonnet gravity couplings. The corresponding boundary action has been also found in \cite{Garousi:2021yyd} in which we are not interested here. The restricted field redefinition at order $\alpha'$ is the following:
\beqa
\delta G^{(1)}_{\mu\nu}\,=\,0\,\,\,;\,\,
  \delta B^{(1)}_{\mu\nu}\,=\,\alpha_1 H_{\mu \nu \alpha} \nabla^{\alpha}\Phi &;&
 \delta\Phi^{(1)}\,=\,\alpha_2 H_{\alpha \beta \gamma} H^{\alpha \beta \gamma} + \alpha_3  \nabla_{\alpha}\Phi\nabla^{\alpha}\Phi\labell{fr3}
\eeqa
where the  coefficients $\alpha_1,\alpha_2,\alpha_3$ are arbitrary parameters. Using this field redefinition it has been shown in \cite{Garousi:2021yyd} that the above action is physically the same as the action proposed by Meissner \cite{Meissner:1996sa}. Using the above field redefinitions and adding total derivative terms, one can write the action \reef{ffinal} in terms of the torsional curvatures as
\beqa
S_1
&=&-\frac{2b_1}{\kappa^2} \alpha'\int d^{26}x \sqrt{-G} e^{-2\Phi}\Bigg[\cR^2_{\rm GB}-  H_{\alpha \beta }{}^{\epsilon }
H_{\gamma \delta \epsilon } \CR^{\alpha \beta \gamma \delta }- \frac{1}{3} H_{\alpha }{}^{\delta \epsilon } H^{\alpha
\beta \gamma } H_{\beta \delta }{}^{\varepsilon } H_{\gamma
\epsilon \varepsilon } \nn\\&&\qquad\qquad+ \frac{1}{36} H_{\alpha \beta \gamma }
H^{\alpha \beta \gamma } H_{\delta \epsilon \varepsilon }
H^{\delta \epsilon \varepsilon } + \frac{1}{3} H_{\gamma
\delta \epsilon } H^{\gamma \delta \epsilon } \CR^{\alpha \beta
}{}_{\alpha \beta }  + \frac{2}{3}
H_{\beta \gamma \delta } H^{\beta \gamma \delta }
\nabla_{\alpha }\Phi \nabla^{\alpha }\Phi\nn\\&&\qquad\qquad + 4 \CR^{\beta \gamma
}{}_{\beta \gamma } \nabla_{\alpha }\Phi \nabla^{\alpha }\Phi
- 16 \CR_{\alpha }{}^{\gamma }{}_{\beta \gamma } \nabla^{\alpha }
\Phi \nabla^{\beta }\Phi \Big]
\eeqa
Note that  the above action contains the couplings \reef{S0bf1} as well as some other couplings.  The above action is background independent, \ie  it is valid for both open and closed spacetime manifolds. However, for closed manifolds one can still use arbitrary field redefinitions to simplify it to the action in \reef{S0bf1}. An alternative way for finding the above action is to first find all independent couplings involving  the torsional Riemann, Ricci and scalar curvatures, as well as the torsion $H$ and $\prt\Phi$ at order $\alpha'$. Then one should reduce them on a circle and impose the $O(1,1)$ symmetry to fix  the parameters of the independent couplings.  %In that way, one finds all parameters are fixed up to an overall factor, \ie no need to impose $O(d,d)$ symmetry on the cosmological reduction of the independent couplings.   
It would be interesting to perform this latter approach to find the couplings at orders $\alpha'^2$ and $\alpha'^3$ for open spacetime manifolds. We expect the torsional Riemann, Ricci and scalar curvatures at order $\alpha'^3$ to appear as in the first term in \reef{Y01} with no restriction on the expansion of $\eps_{8}\eps_{8}\cR^4$.

We have seen that using field redefinitions, one can write the spacetime effective actions in terms of the torsional curvatures. One may ask if it is possible to write the world-volume couplings of D-brane/O-plane effective actions in terms of the torsional curvatures as well?   Because the field redefinitions are  used for the spacetime effective actions to be written in terms of the torsional curvatures, one is not allowed to use another field redefinitions for the world-volume couplings. However, since these objects are considered as probe, one can impose the equations of motion in their world-volume effective actions \cite{Robbins:2014ara}.  Hence, there is unique form for the world-volume couplings up to spacetime  equations of motion, the  world-volume total derivative terms and the Bianchi identities. The world-volume couplings found in the literature  are not in terms of the torsional curvatures. For example, the world-volume couplings at order $\alpha'^2$ in the superstring theory which are produced by the contact terms of the disk-level S-matrix element of two NS-NS vertex operators or one NS-NS and one R-R vertex operators at order $\alpha'^2$, are found in \cite{Bachas:1999um,Garousi:2009dj,Garousi:2010ki}. %They can not be written in terms of the torsional Riemann curvatures. In fact
The curvature terms in these couplings have even number of transverse indices whereas the $\nabla H$-terms have odd number of transverse indices. Using the relation \reef{nonlinear2}, the Riemann curvature $R_{\mu\nu\alpha\beta}$ can be written as $\cR_{\mu\nu\alpha\beta}+\cR_{\alpha\beta\mu\nu}$ plus $HH$-terms   and $\nabla_{[\mu} H_{\nu]\alpha\beta}$ can be written  as $\cR_{\mu\nu\alpha\beta}-\cR_{\alpha\beta\mu\nu}$
 Hence, the curvature terms can be written in terms of the torsional Riemann curvature that has even number of transverse indices, and  the $\nabla H$-terms can be written in terms of the torsional Riemann curvature that has odd number of transverse indices. In fact, any expression in terms of Riemann curvatures $\nabla H$ and  $H$ can be rewritten in terms of the torsional Riemann curvature and $H$. As we have seen in this paper, the couplings in terms of the  torsional Riemann curvature and $H$ have simpler form.  %However, in these studies, the standard propagators are assumed for the massless open string gauge field and the transverse scalar fields. Using such standard propagators, the leading $\alpha'$ order of the disk-level S-matrix element are reproduced completely by the DBI action. On the other hand, it has been observed in \cite{Hosseini:2022vrr}  that the background independence of the world-volume couplings dictates that the corresponding propagators in the bosonic string theory are not the standard propagators. The non-standard propagators produce some extra contact terms \cite{Hosseini:2022vrr}. It would be interesting to examine in details whether the background independence of the world-volume couplings dictates   the propagators in the superstring theory are the standard propagators or not. If they are not standard ones, then there might be some extra world-volume couplings that must be added to the couplings found in \cite{Bachas:1999um,Garousi:2009dj,Garousi:2010ki}. The result might then be in terms of the torsional Riemann curvature.

%S-duality in the presence of boundary

 \vskip .3 cm
{\bf Acknowledgments}:   I would like to thank Linus Wulff for useful conversations. % under grant  1/50251(1398/04/31).

%\newpage

\end{document}